\documentclass[a4,prb,onecolumn,amsfonts,amssymb]{revtex4-1}
\usepackage{amsmath}
\usepackage{graphicx}
\usepackage{color}
\usepackage{hyperref}
\usepackage{psfrag}
\usepackage{stmaryrd}
\usepackage{amssymb}
\usepackage{wasysym}
\usepackage[utf8]{inputenc}
\usepackage{hyperref}

\def \degre {$^\mathrm{o}$}

\begin{document}

\title{Experimental evidence of a triadic resonance of plane inertial waves in a rotating fluid}

\author{Guilhem Bordes$^1$}
\email{guilhem.bordes@ens-lyon.fr}
\author{Fr\'{e}d\'{e}ric Moisy$^2$}
\email{moisy@fast.u-psud.fr}
\author{Thierry Dauxois$^1$}
\email{thierry.dauxois@ens-lyon.fr}
\author{Pierre-Philippe Cortet$^2$}
\email{ppcortet@fast.u-psud.fr} \affiliation{$^1$Laboratoire de
Physique de l'\'{E}cole Normale Sup\'{e}rieure de Lyon, CNRS and
Universit\'{e} de Lyon, 46 All\'{e}e d'Italie, 69007 Lyon, France}
\affiliation{$^2$Laboratoire FAST, CNRS, Univ Paris-Sud, UPMC Univ
Paris 06, France}

\date{\today}

\begin{abstract}

Plane inertial waves are generated using a wavemaker, made of
oscillating stacked plates, in a rotating water tank. Using
particle image velocimetry, we observe that, after a transient,
the primary plane wave is subject to a subharmonic instability and
excites two secondary plane waves. The measured frequencies and
wavevectors of these secondary waves are in quantitative agreement
with the predictions of the triadic resonance mechanism. The
secondary wavevectors are found systematically more normal to the
rotation axis than the primary wavevector: this feature
illustrates the basic mechanism at the origin of the energy
transfers towards slow, quasi two-dimensional, motions in rotating
turbulence.

\end{abstract}

\maketitle

\section{Introduction}

Rotating and stratified fluids support the existence of two
classes of anisotropic dispersive waves,  called respectively
inertial and internal waves, which play a major role in the
dynamics of astrophysical and geophysical
flows.~\cite{Greenspan1968,Lighthill1978,Pedlosky1987} These waves
share a number of similar properties, such as a group velocity
normal to the phase velocity. Remarkably, in both cases, the frequency
of the wave selects only its direction of propagation, whereas the
wavelength is selected by other physical properties of the
system, such as the boundary
conditions or the viscosity.~\cite{Lighthill1978,Phillips1963,Gostiaux2006}

Most of the previous laboratory experiments on inertial waves in
rotating fluids have focused on inertial modes or wave attractors
in closed
containers,~\cite{Fultz1959,McEwan1970,Manasseh1994,Maas2001,Duguet2005,Duguet2006,Meunier2008}
whereas less attention has been paid to propagative inertial wave
beams. Inertial modes and attractors are generated either from a
disturbance of significant size compared to the
container,~\cite{Fultz1959} or more classically from global
forcing.~\cite{McEwan1970,Manasseh1994,Maas2001,Duguet2005,Duguet2006,Meunier2008}
Inertial modes are also detected in the ensemble average of
rotating turbulence experiments in closed
containers.~\cite{Bewley2007,Lamriben2011a} On the other hand,
localized propagative inertial wave beams have been investigated
recently in experiments using particle image velocimetry
(PIV).~\cite{Messio2008,Cortet2010}

A monochromatic internal or inertial wave of finite amplitude may
become unstable with respect to a parametric subharmonic
instability.~\cite{Thorpe1969,McEwan1971,Benielli1998,Staquet2002}
This instability originates from a nonlinear resonant interaction
of three waves, and induces an energy transfer from the primary
wave towards two secondary waves of lower frequencies. This
instability has received considerable interest in the case of
internal gravity waves,~\cite{Staquet2002} because it is believed
to provide an efficient mechanism of dissipation in the oceans, by
allowing a transfer of energy from the large to the small
scales.~\cite{Olbers1981,Kunze2004,MacKinnon2005}

Parametric instability is a generic mechanism expected for any
forced oscillator. A pendulum forced at twice its natural
frequency provides a classical illustration of this mechanism.
Here, the ``parameter'' is the natural frequency of the pendulum,
which is modulated in time through variations of the gravity or
pendulum length. Weakly nonlinear theory shows that the energy of
the excitation, at frequency $\sigma_0$, is transferred to the
pendulum at its natural frequency $\sigma_0/2$, resulting in an
exponential growth of the oscillation amplitude.

In the case of inertial (resp. internal) waves, the ``parameter''
is now the so-called Coriolis frequency $f = 2\Omega$, with
$\Omega$ the rotation rate (resp. the Brunt-V\"{a}is\"{a}l\"{a}
frequency $N$). In the presence of a primary wave of frequency
$\sigma_0$, this ``parameter'' becomes locally modulated in time
at frequency $\sigma_0$, and is hence able to excite secondary
waves of lower natural frequency. However, here a continuum of
frequencies can be excited, so that the frequencies $\sigma_1$ and
$\sigma_2$ of the secondary waves are not necessarily half the
excitation frequency, but they nevertheless have to satisfy the
resonant condition $\sigma_1+\sigma_2=\sigma_0$. Interestingly, in
the absence of dissipation, the standard pendulum-like resonance
$\sigma_1 = \sigma_2 = \sigma_0/2$ is recovered both for inertial
and internal waves, and the corresponding secondary waves have
vanishing wavelengths.~\cite{Staquet2002} Viscosity is responsible
here for the lift of degeneracy, by selecting a maximum growth
rate corresponding to finite wavelengths, with frequencies
$\sigma_1$ and $\sigma_2$ splitted on both sides of
$\sigma_0/2$.~\cite{Koudella2006}

The parametric subharmonic instability has been investigated in
detail for internal gravity waves.~\cite{Staquet2002,Koudella2006}
On the other hand, this instability mechanism has received less
attention in the case of pure inertial waves (i.e., in absence of
stratification), probably because of the lower importance of
rotation effects compared to stratification effects in most
geophysical flows. It has been observed in numerical simulations
of inertial modes in a periodically compressed rotating
cylinder.~\cite{Duguet2005,Duguet2006} To our knowledge,
parametric instability in the simpler geometry of plane inertial
waves has not been investigated so far, and is the subject of this
paper. A fundamental motivation for this work is the key role
played by triadic interactions of inertial waves in the problem of
the generation of slow quasi-2D flows in rotating
turbulence.~\cite{Smith1999,Cambon2008,Lamriben2011b} The
parametric subharmonic instability indeed provides a simple but
nontrivial mechanism for anisotropic energy transfers from modes
of arbitrary wavevectors towards lower frequency modes, of
wavevector closer to the plane normal to the rotation axis (i.e.,
more ``horizontal'' by convention). Note that this nonlinear
mechanism may however be in competition with a linear mechanism
---the radiation of inertial waves along the rotation axis--- which
has also been shown to support the formation of vertical columnar
structures.\cite{Staplehurst2008} The relative importance of these
two mechanisms is governed by the Rossby number, defined as $Ro =
(\tau_{\rm nl} \Omega)^{-1}$, with $\Omega^{-1}$ the linear
timescale and $\tau_{\rm nl} = L/U$ the nonlinear timescale based
on the characteristic velocity $U$ and length scale $L$. In
rotating turbulence with $Ro \ll 1$, the anisotropy growth should
hence be dominated by the nonlinear triadic interactions, whereas
for $Ro = O(1)$ both mechanisms should be at play.

In this paper, we report the first experimental observation of the
destabilization of a primary plane inertial wave and the
subsequent excitation of subharmonic secondary waves. To produce a
plane inertial wave of sufficient spatial extent, and hence of
well-defined wavevector ${\bf k}_0$,  we have made use of a wave
generator already developed for internal waves in stratified
fluids.\cite{Gostiaux2007,Mercier2008,Mercier2010} Wave beams of
tunable shape and orientation can be generated with this
wavemaker. We show that, after a transient, the excited plane wave
undergoes a parametric subharmonic instability. This instability
leads to the excitation of two secondary plane waves, with
wavevectors which are systematically more ``horizontal'' than the
primary wavevector. We show that the predictions from the resonant
triadic interaction theory for inertial waves, as described by
Smith and Waleffe,\cite{Smith1999} are in excellent agreement with
our experimental results. In particular, the frequencies and
wavenumbers of the secondary waves accurately match the expected
theoretical values.

\section{Inertial plane wave generation}

\subsection{Structure of a plane inertial wave}

We first briefly recall the main properties of inertial waves in a
homogeneous fluid rotating at a constant rate~$\Omega$. In the
rotating frame, the restoring nature of the Coriolis force is
responsible for the propagation of the inertial waves, for
frequencies $\sigma \leq f$, where $f=2\Omega$ is the Coriolis
parameter. Fluid particles excited at frequency~$\sigma$ describe
anticyclonic circles in a plane tilted at an angle
$\theta=\cos^{-1}(\sigma/f)$ with respect to the horizontal, and
the phase of this circular motion propagates perpendicularly to
this tilted plane.

\begin{figure}
    \centerline{\includegraphics[width=8.5cm]{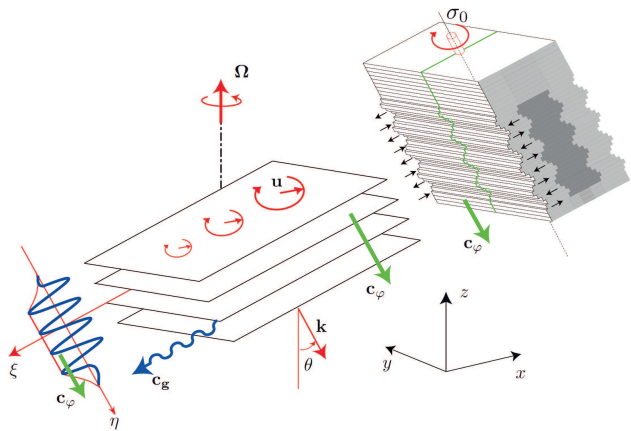}}
    \caption{(Color online) Schematic representation of the wave
    generator. The excited plane inertial wave has a frequency
    $\sigma_0$, a downward phase velocity, a negative helicity ($s_0 =
    -1$), and propagates at an angle $\theta = \cos^{-1}(\sigma_0 /
    f)$, with $f=2\Omega$ the Coriolis parameter.\label{wavemaker}}
\end{figure}

The equations of motion for a viscous fluid in a frame rotating at
a rate $\Omega = f/2$ around the axis $z$ are
\begin{eqnarray}
\partial_t {\bf u} +  ({\bf u} \cdot \nabla) {\bf u}
 &=& -\frac{1}{\rho}{\bf
\nabla}p - f {\bf e}_z \times {\bf u} +\nu\nabla^2{\bf u}, \label{Navier_Stokes_eq}\\
{\bf\nabla}\bf\cdotp{\bf u} &=& 0, \label{eq_div_0}
\end{eqnarray}
where ${\bf u}=(u_x,u_y,u_z)$ is the velocity field in cartesian
coordinates ${\bf x}=(x,y,z)$. In the following, we restrict to
the case of a flow invariant along the horizontal direction $y$.
The fluid being incompressible, the motion in the vertical plane
$(x,z)$ may be described by a streamfunction $\psi(x,z)$, such
that ${\bf u}=(\partial_z \psi, u_y, -\partial_x \psi)$.
Neglecting viscosity, the linearized equations for small velocity
disturbances are
\begin{eqnarray}
\partial_t \partial_z \psi & = & -\frac{1}{\rho} \partial_x p + f u_y, \\
\partial_t u_y  & = & - f \partial_z \psi, \\
- \partial_t \partial_x \psi & = & -\frac{1}{\rho} \partial_z p.
\end{eqnarray}
These equations may be combined to obtain the equation of
propagation for inertial waves,
\begin{eqnarray}
\partial_{tt} (\partial_{xx} + \partial_{zz}) \psi + f^2 \partial_{zz} \psi &=& 0. \label{INWeq}
\end{eqnarray}
Considering a plane wave solution of frequency $\sigma$ and
wavevector ${\bf k}=(k,0,m)$,
\begin{eqnarray}
\psi(x,z,t) = \psi_0\,e^{i({\bf k}\cdot{\bf x}-\sigma t)} + {\rm
c.c.}
\end{eqnarray}
(where c.c. means complex conjugate), we obtain the anisotropic
dispersion relation for inertial waves
\begin{eqnarray}
 \sigma &=& s f\frac{m}{\kappa} = s f\cos\theta, \label{disp_eq}
\end{eqnarray}
with $\kappa=(k^2+m^2)^{1/2}$, $s=\pm 1$, and $\theta$ the angle
between ${\bf k}$ and the rotation axis (see
Fig.~\ref{wavemaker}). We see from Eq.~(\ref{disp_eq}) that
a given frequency $\sigma$ lower than $f$ selects a propagation angle
$\pm \theta$, without specifying the norm of the wavevector $\kappa$.
The corresponding velocity field is given by
\begin{eqnarray}
u_x & = & i m \, \psi_0 \,e^{i(kx+mz -\sigma t)} + {\rm
c.c.} \label{u_component}\\
u_y & = & s \kappa \, \psi_0 \,e^{i(kx+mz -\sigma t)} + {\rm
c.c.} \label{v_component}\\
u_z & = & - i k \, \psi_0 \,e^{i(kx+mz -\sigma t)} + {\rm c.c.}
\label{w_component}
\end{eqnarray}
We recover here that the fluid particles describe anticyclonic
circular motions in tilted planes perpendicular to ${\bf k}$, as
sketched in Fig.~\ref{wavemaker}. The wave travels with a phase
velocity ${\bf c}_\varphi = \sigma {\bf k} / \kappa^2$ and a group
velocity ${\bf c}_g = \nabla_{\bf k} \sigma$ normal to ${\bf
c}_\varphi$. The vorticity $\boldsymbol{\omega} = \nabla \times {\bf u}$, given
by
\begin{equation}\label{waleffe2ter}
\boldsymbol{\omega} = - s \kappa \mathbf{u},
\end{equation}
is associated to the shearing motion between planes of constant
phase. Because the velocity and vorticity are aligned, inertial
waves are also called {\it helical waves}, and the sign $s$ in
Eq.~(\ref{disp_eq}) identifies to the sign of the wave helicity
${\bf u} \cdot \boldsymbol{\omega}$, with $s=+1$ for a
right-handed wave and $s=-1$ for a left-handed wave. For instance,
in the classical St.\,Andrew's wave pattern emitted from a point
source,\cite{Cortet2010} the two upper beams are right-handed and
the two lower beams are left-handed, although the fluid motion is
always anticyclonic.

\subsection{Generation of a plane inertial wave}

In order to generate a plane inertial wave, we have made use of a
wavemaker, introduced by Gostiaux {\it et al.},\cite{Gostiaux2007}
which was originally designed to generate internal gravity waves
(see Mercier {\it et al.}\cite{Mercier2010} for a detailed
characterization of the wavemaker). This wavemaker consists in a
series of oscillating stacked plates, designed to reproduce the
fluid motion in the bulk of an internal gravity wave invariant
along $y$. The use of this internal wave generator for the
generation of inertial waves is motivated by the similarity of the
spatial structure of the two types of waves in the vertical plane
$(x,z)$. However, the fluid motion in the internal wave is a
simple oscillating translation in the direction of the group
velocity, whereas fluid particles describe anticyclonic circular
translation in the case of inertial waves. As a consequence, the
oscillating plates of the wavemaker only force the longitudinal
component of the circular motion of the inertial waves, whereas
the lateral component is let to freely adjust according to the
spatial structure of the wave solution.

The wavemaker is made of a series of 48 parallelepipedic plates
stacked around a helical camshaft, with the appropriate shifts
between successive cames in order to form a sinusoidal profile at
the surface of the generator. We introduce  the local coordinate
system $(\xi,y,\eta)$, tilted at an angle $\theta$ about $y$,
where $\xi$ is along the wave propagation and $\eta$ is parallel
to the camshaft axis (see Fig.~\ref{wavemaker}). The group
velocity and the phase velocity of the wave are oriented along
$\xi$ and $\eta$ respectively. As the camshaft rotates at
frequency $\sigma_0$, the plates, which are constrained in the $y$
direction, oscillate back and forth along $\xi$. The sign of the
rotation of the helical camshaft selects the helicity of the
excited wave, and hence an upward or downward phase velocity. In
the present experiment, the rotation of the camshaft is set to
produce a downward phase velocity,  resulting in a left-handed
inertial wave of negative helicity $s_0 = -1$.

The cames are 14~cm wide in the $y$ direction, and their
eccentricities are chosen to produce a sinusoidal displacement
profile, $\xi_0(\eta)=\xi_o\,\sin(\kappa_0 \eta)$, of wavelength
$\lambda=2\pi/\kappa_0 = 7.6$~cm and amplitude $\xi_o=0.5$~cm at
the center of the beam. The wave beam has a width 30.5~cm with a
smooth decrease to 0 at the borders, and contains approximately 4
wavelengths. The generator is only forcing the $\xi$ component of
the inertial wave, and the $y$ component is found to adjust
according to the inertial wave structure after a distance of order
of 2~cm.

The wavemaker is placed in a tank of 120~cm length, 80~cm width
and 70~cm depth which is filled with 58~cm of water. The tank is
mounted on the precision rotating platform ``Gyroflow'' of 2~m in
diameter. The angular velocity $\Omega$ of the platform is set in
the range 1.05 to 3.15~rad~s$^{-1}$, with relative fluctuations
$\Delta \Omega / \Omega$ less than $10^{-3}$. A cover is placed at
the free surface, preventing from disturbances due to residual
surface waves. The rotation of the fluid is set long before each
experiment (at least 1 hour) in order to avoid transient spin-up
recirculations and to achieve a clean solid body rotation.

The propagation angle $\theta$ of the inertial wave is varied by
changing the rotation rate of the platform, while keeping the
wavemaker frequency constant, $\sigma_0=1.05$~rad~s$^{-1}$. This
allows to  have a fixed wave amplitude
$\sigma_0\,\xi_o=0.52$~cm~s$^{-1}$ for all angles. The Coriolis
parameter has been varied in the range $f=1.004\,\sigma_0$ to
$3\,\sigma_0$, corresponding to angles $\theta$ from 5\degre~to
70\degre. For each value of the rotation rate, the axis of the
wavemaker camshaft is tilted to the corresponding angle $\theta =
\cos^{-1} (\sigma_0 / f)$, in order to keep the plate oscillation
aligned with the fluid motion in the excited wave. As a
consequence, the efficiency of the forcing should not depend
significantly on the angle $\theta$. For each experiment, the
fluid is first reset to a solid body rotation before the wavemaker
is started.

\subsection{PIV measurements}

Velocity fields are measured using a 2D particle image velocimetry
(PIV) system~\cite{Davis,pivmat} mounted on the rotating platform.
The flow is seeded by 10~$\mu$m tracer particles, and illuminated
by a vertical laser sheet, generated by a 140~mJ Nd:YAG pulsed
laser. A vertical 59$\times$59 cm$^2$ field of view is acquired by
a 14 bits 2048$\times$2048 pixels camera synchronized with the
laser pulses. For each rotation rate, a set of 3200 images is
recorded, at a frequency of 4~Hz, representing 24 images per
wavemaker period. This frame rate is set to achieve a typical
particle displacement of 5 to 10 pixels between each frame,
ensuring an optimal signal-to-noise ratio for the velocity
measurement. PIV computations are performed over successive
images, on 32$\times$32 pixels interrogation windows with 50\%
overlap. The spatial resolution is 4.6~mm, which represents 17
points per wavelength of the inertial wave.

\begin{figure}
    \centerline{\includegraphics[width=8.5cm]{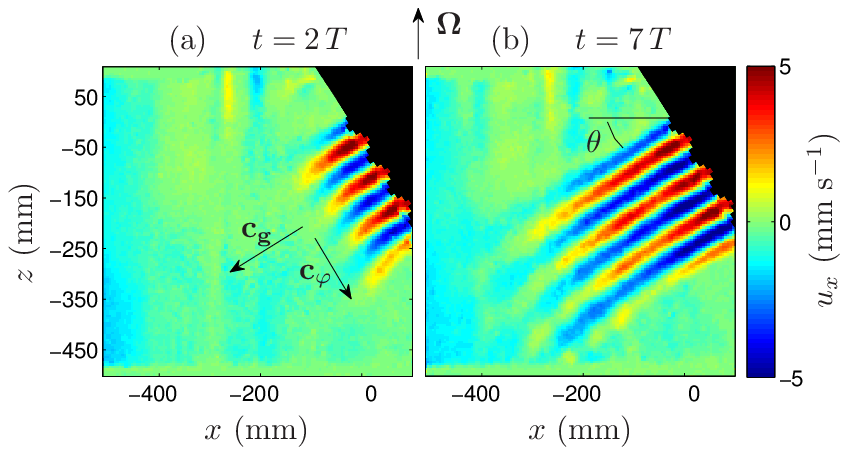}}
    \caption{(Color) Horizontal velocity field after 2 and 7 periods from the
    start of the wavemaker for $\sigma_0/f=0.84$. The wavemaker is on
    the top-right, forcing a wave propagating along ${\bf c_g}$ with a
    phase propagating along ${\bf c_\varphi}$.\label{propag_time}}
\end{figure}

Figure~\ref{propag_time} shows typical instantaneous horizontal
velocity fields after 2 and 7 periods $T=2\pi/\sigma_0$ from the
start of generator, for an experiment performed with
$\sigma_0/f=0.84$. A well defined truncated plane wave propagates
downward, making an angle $\theta=\cos^{-1}(\sigma_0/f) \simeq
34$\degre~ to the horizontal. The front of the plane wave is
propagating at a velocity $8.3\pm0.6$~mm~s$^{-1}$, which agrees
well with the expected group velocity $c_g= f \sin\theta/\kappa =
8.5$~mm~s$^{-1}$. The phase velocity is downward, normal to the
group velocity, and also agrees with the expected value
$c_{\varphi}=\sigma_0/\kappa=12.7$~mm~s$^{-1}$.

Two sources of noise have been identified, which can be seen in
the temporal energy spectrum of the velocity fields
(Fig.~\ref{spectre}, described in the next subsection): an
oscillatory motion at frequency $\sigma=\Omega=0.5 f$, due to a
residual modulation of the rotation rate of the platform, and
slowly drifting thermal convection structures at frequency $\sigma
\rightarrow 0$, due to slight temperature inhomogeneities in the
tank. Both effects contribute to a velocity noise of order of
0.2~mm~s$^{-1}$, i.e. 25 times lower than the wave amplitude close
to the wavemaker. This noise could be safely removed using a
temporal Fourier filtering of the velocity fields at the forcing
frequency $\sigma_0$. This filtering however fails in the
particular case where $\sigma_0=\Omega$, for which the mechanical
noise of the platform cannot be filtered out of the inertial wave
signal.

The wavemaker is found to successfully generate well defined plane
waves for frequencies $\sigma_0 \geq 0.65 f$. For lower frequency,
i.e. for steeper angle of propagation
[$\theta=\cos^{-1}(\sigma_0/f)>50$\degre], the wave pattern shows
significant departure from the expected plane wave profile, which
may be attributed to the interference of the incident wave with
the reflected wave on the bottom of the tank.

\section{Subharmonic Instability}

\subsection{Experimental observations}

\begin{figure}
    \centerline{\includegraphics[width=7cm]{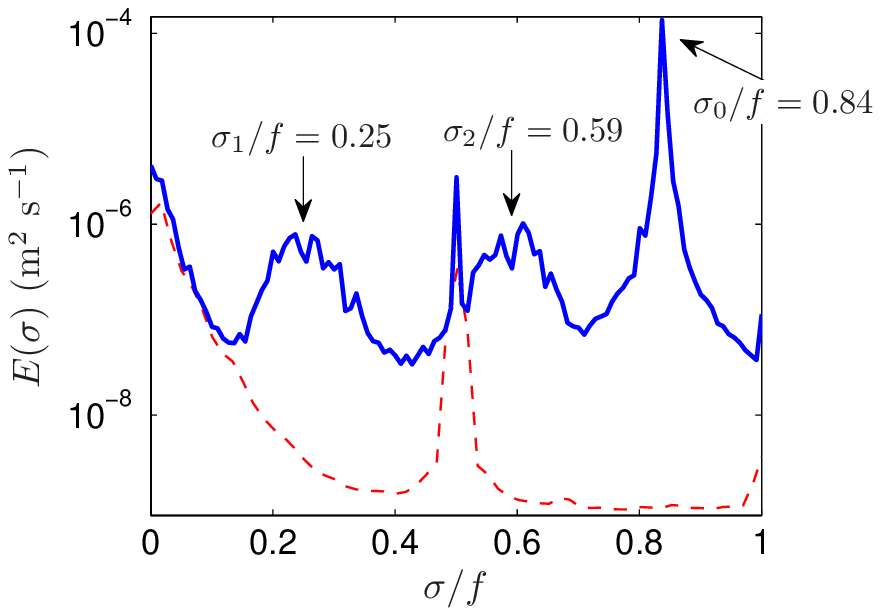}}
    \caption{(Color online) Temporal energy spectra for two
    experiments performed at rotation rate $\Omega=0.63$~rad~s$^{-1}$
    with (continuous line) and without (dashed line) the wavemaker
    operating at $\sigma_0/f=0.84$. The spectrum with the generator
    working has been computed on the time interval between 24 and 116
    periods after the start of the generator. The peak at
    $\sigma/f=0.5$ present in the two spectra is the trace of the
    mechanical noise of the platform at the rotation frequency
    $\sigma=\Omega$, whereas the low frequencies are due to thermal
    convection effects (see text). \label{spectre}}
\end{figure}

After a few excitation periods, the front of the inertial wave has
travelled outside the region of interest, and the inertial wave
can be considered locally in a stationary regime. However, after
typically 15 wavemaker periods (the exact value depends on the
ratio $\sigma_0/f$), the inertial wave becomes unstable and show
slow disturbances of scale slightly smaller than the excited
wavelength.

We have characterized this instability using Fourier analysis of
the PIV time series. We compute, at each location $(x,z)$ of the
PIV field, the temporal Fourier transform of the two velocity
components over a temporal window $\Delta t$,
\begin{equation}\label{usigma}
\hat {\bf u}_\sigma (x,z) = \frac{1}{\sqrt{2\pi}}
\int_{t_0}^{t_0+\Delta t} {\bf u}(x,z,t) e^{i \sigma t} dt.
\end{equation}
The temporal energy spectrum is then defined as
\begin{equation}\label{fourier}
E(\sigma) = \langle | \hat {\bf u}_\sigma |^2 \rangle_{x,z},
\end{equation}
where  $\langle \cdot \rangle_{x,z}$ is the spatial average over
the PIV field.

If we compute $E(\sigma)$ over a temporal window $\Delta t$
spanning a few excitation periods, we observe, as $t_0$ is
increased, the emergence of two broad peaks at frequencies smaller
than the excitation frequency $\sigma_0$, suggesting the growth of
a subharmonic instability. These two subharmonic peaks can be seen
in Fig.~\ref{spectre}, for an experiment performed at rotation
rate $\Omega=0.63$~rad~s$^{-1}$ with the wavemaker operating at
$\sigma_0/f=0.84$. Here, the temporal window $\Delta t$ is chosen
equal to 92 wavemaker periods, yielding a spectral resolution of
$\Delta \sigma = 2\pi / \Delta t \simeq 9 \times 10^{-3} \, f$.
The two secondary peaks are centered on $\sigma_1/f=0.25 \pm 0.03$
and $\sigma_2/f=0.59\pm 0.03$, and their sum matches well with the
forcing frequency $\sigma_0/f = 0.84$, as expected for a
subharmonic resonance. The significant width of the secondary
peaks, of order $0.07\,f$, indicates that this resonance is weakly
selective. This broad-band selection will be further discussed in
Sec.~\ref{sec:smuw}.

The subharmonic instability of the primary wave is found for all
forcing frequencies $\sigma_0$ ranging from $0.65f$ to $f$; the
measured frequencies $\sigma_{1,2}$ are given in Tab.~\ref{tab:1}.
The absence of clear subharmonic instability at lower forcing
frequency may be due to an intrinsic stability of the primary wave
for $\sigma_0 < 0.65f$, or to the low quality of the plane wave at
steep angles because of the interference with the reflected wave
beam on the bottom of the tank.

\begin{table}
\begin{tabular}{cccc}
\hline \hline
~~~~$\sigma_0/f$~~~~ & ~~~~$(\sigma_1+\sigma_2)/f$~~~~ & ~~~~$\sigma_1/f$~~~~ & ~~~~$\sigma_2/f$~~~~\\
\hline
0.64 & 0.64 & 0.19 & 0.45 \\
0.71 & 0.71 & 0.21 & 0.50 \\
0.84 & 0.84 & 0.25 & 0.59 \\
0.91 & 0.94 & 0.27 & 0.67 \\
0.95 & 0.97 & 0.29 & 0.68 \\
0.98 & 0.98 & 0.32 & 0.66 \\
0.99 & 1.00 & 0.34 & 0.66 \\
\hline \hline
\end{tabular}
\caption{Frequencies of the secondary waves $\sigma_1/f$ and $\sigma_2/f$,
determined from the peaks in the temporal
energy spectra, as a function of the frequency of the primary wave
$\sigma_0/f$. The uncertainty for $\sigma_1/f$ and $\sigma_2/f$ is $\pm 0.03$.\label{tab:1}}
\end{table}

Using temporal Hilbert filtering,~\cite{Mercier2008,Croquette1989}
the spatial structure of the wave amplitude ${\bf u}_o({\bf x})$
and phase $\varphi({\bf x},t)={\bf k}\cdot{\bf x}-\sigma t$ can be
extracted for each secondary wave. The procedure consists in (i)
computing the Fourier transform $\hat {\bf u}_\sigma (x,z)$ of the
velocity field according to Eq.~(\ref{usigma}), with a temporal
window $\Delta t$ of at least 42 excitation periods; (ii)
band-pass filtering $\hat {\bf u}_\sigma (x,z)$ around the
frequency of interest $\sigma_{1}$ or $\sigma_{2}$ with a
bandwidth of $\delta\sigma=2.0\,10^{-2}f$, but without including
the associated negative frequency; (iii) reconstructing the
complex velocity field by computing the inverse Fourier transform
(including a factor 2, which accounts for the redundant negative
frequency, in order to conserve energy),
\begin{equation}
{\bf u}_H({\bf x},t)={\bf u}_o({\bf x})\,e^{i \varphi({\bf
x},t)}.
\end{equation}
The physical velocity field is finally given by
\textit{Re}$\left({\bf u}_H \right)$. The wave amplitude ${\bf
u}_o$ and phase field $\varphi$ are finally obtained from the
Hilbert-filtered field ${\bf u}_H$.

\begin{figure}
    \centerline{\includegraphics[width=13cm]{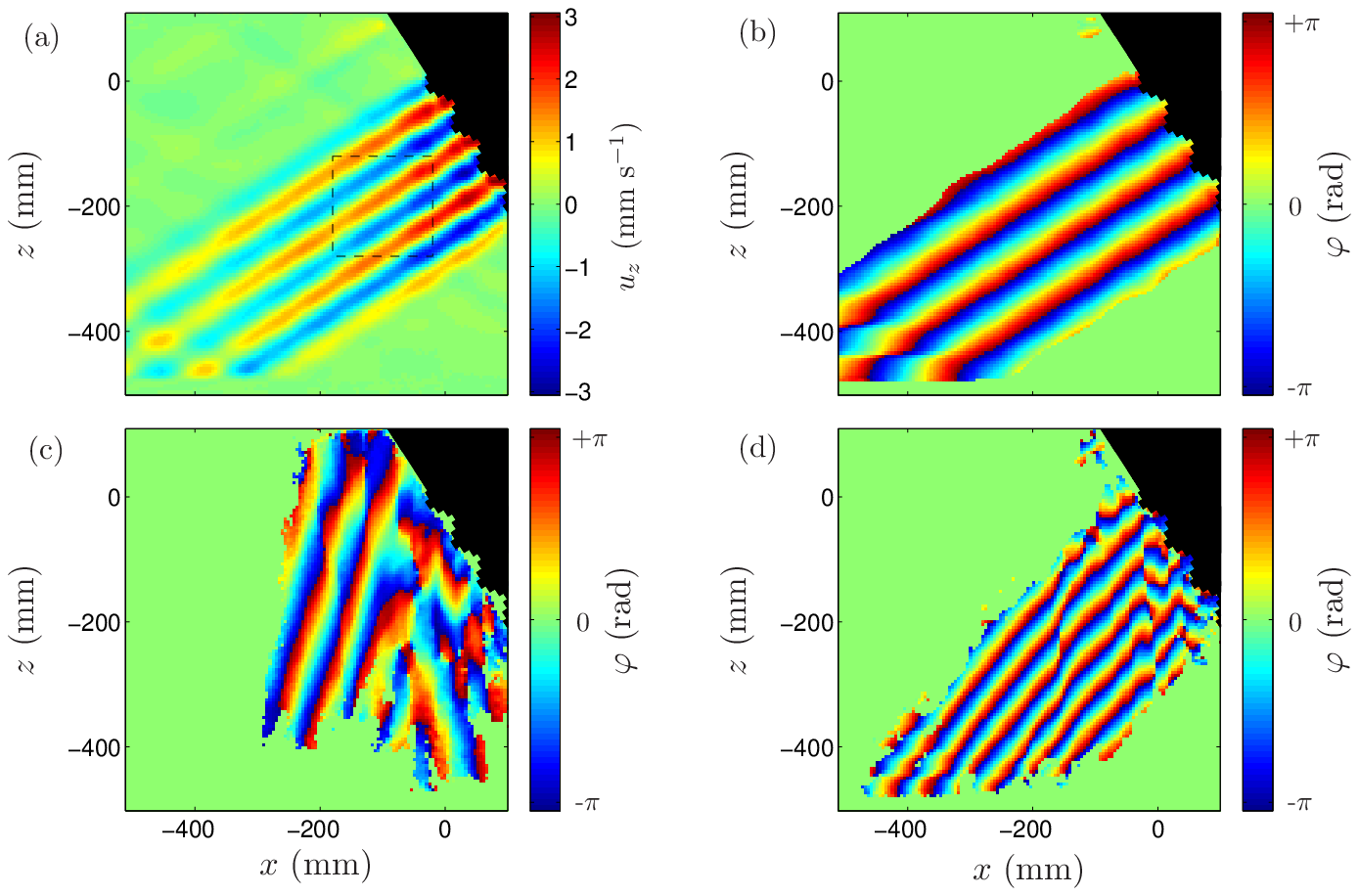}}
    \caption{(Color) Hilbert filtered vertical velocity (a) and phase
    (b) of the primary wave at $\sigma_0/f=0.84$, and phase of the
    Hilbert filtered first [(c), $\sigma_1/f=0.25$] and second [(d),
    $\sigma_2/f=0.59$] subharmonic waves. The phase is displayed only
    where the wave amplitude is larger than $1.3\,10^{-1}\,\sigma_0
    \xi_0$ for (b) and $7.7\,10^{-3}\,\sigma_0 \xi_0$ for (c) and (d).
    In (a), the square in dashed lines indicates the region where the
    primary wave amplitude $A_0$ has been measured.\label{sub12}}
\end{figure}

In Figs.~\ref{sub12}(c) and (d), for the experiment at
$\sigma_0/f=0.84$, we show the maps of the phase of the secondary
waves, extracted from Hilbert filtering at frequencies $\sigma_1$
and $\sigma_2$ respectively. It is worth to note, as can be
verified from Fig.~\ref{spectre}, that the corresponding typical
velocity amplitude is at least ten times smaller than for the
primary wave [see Fig.~\ref{sub12}(a)]. The spatial structures of
the phase of these secondary waves are not as clearly defined as
for the primary wave [Fig.~\ref{sub12}(b)]. In particular,
dislocations can be distinguished in the phase field. The finite
extent of the primary wave and its spatial decay due to viscous
attenuation are probably responsible for this departure of the
secondary waves from pure plane waves. It is also important to
note that the monochromaticity of the first subharmonic wave
[Fig.~\ref{sub12}(c)] is affected by interferences with its
reflection on the wavemaker which is due to the fact this
secondary wave is propagating toward the wavemaker. However, to a
reasonable degree of accuracy, the two secondary waves can be
considered locally as plane waves, characterized by local
wavevectors ${\bf k}_1$ and ${\bf k}_2$.

\subsection{Helical modes}

The approximate plane wave structure of the two secondary waves
suggests to analyze the instability in terms of a triadic
resonance between the primary wave, of wavevector ${\bf k}_0$,
and the two secondary waves, of wavevectors ${\bf k}_1$
and ${\bf k}_2$. This resonance may be conveniently analyzed in
the framework of the helical decomposition, introduced by
Waleffe,~\cite{Waleffe1992,Waleffe1993} which we briefly
recall here.

Helical modes have been introduced as a general spectral
decomposition basis, which is useful to analyze the energy
transfers via triadic interactions. Although this decomposition
also applies for non-rotating flows, it is particularly relevant
for rotating flows, because inertial plane waves have exactly the
structure of helical modes.~\cite{Waleffe1993} Any velocity field
can actually be decomposed as a superposition of helical modes of
amplitudes $A_{s_{\bf k}}({\bf k},t)$,
\begin{eqnarray}\label{waleffe1}
{\bf u}({\bf x},t)=\sum_{\bf k}\sum_{s_{\bf k}=\pm 1} A_{s_{\bf
k}}({\bf k},t)\,{\bf h}_{s_{\bf k}}({\bf k})e^{i({\bf k\cdot
x}-\sigma_{s_{\bf k}}^{\bf k} t)},
\end{eqnarray}
where $\sigma_{s_{\bf k}}^{\bf k}$ is the frequency associated to
a plane wave of wavevector ${\bf k}$ and helicity sign $s_{\bf
k}$. The helical mode ${\bf h}_{s_{\bf k}}({\bf k})$ is normal to
${\bf k}$ (by incompressibility), and given by
\begin{equation}\label{waleffe2}
{\bf h}_{s_{\bf k}}({\bf k})=\frac{\mathbf{k}}{|\mathbf{k}|} \times
\frac{\mathbf{k}\times \mathbf{e_z}}{|\mathbf{k}\times
\mathbf{e_z}|}+i s_{\bf k} \frac{\mathbf{k}\times
\mathbf{e_z}}{|\mathbf{k}\times \mathbf{e_z}|},
\end{equation}
where $s_{\bf k}=\pm 1$ is the sign of the mode
helicity.\cite{fn_sign} Injecting the
decomposition~(\ref{waleffe1}) into the Navier-Stokes
equation~(\ref{Navier_Stokes_eq}) yields
\begin{eqnarray}\label{NavierStokes_waleffe}
\left(\frac{\partial}{\partial t} +\nu\kappa^2 \right)A_{\bf
k} = \frac{1}{2}\sum C_{\bf kpq}^{s_{\bf k}s_{\bf p}s_{\bf q}}A^*_{\bf
p}A^*_{\bf q}e^{i(\sigma_{\bf k}+\sigma_{\bf p}+\sigma_{\bf
q})t},
\end{eqnarray}
with stars denoting complex conjugate, and $A_{\bf k}$,
$\sigma_{\bf k}$ being short-hands for $A_{s_{\bf k}}({\bf k},t)$,
$\sigma_{s_{\bf k}}^{\bf k}$. In Eq.~(\ref{NavierStokes_waleffe}),
the sum is to be understood over all wavevectors ${\bf p}$ and
${\bf q}$ such that ${\bf k+p+q=0}$ and all corresponding helicity
signs $s_{\bf p}$ and~$s_{\bf q}$. In the following, the equation
${\bf k+p+q=0}$ will be referred to as the spatial resonance
condition for a triad of helical modes. The interaction
coefficient is given by
\begin{eqnarray}\label{waleffe5}
C_{\bf kpq}^{s_{\bf k}s_{\bf p}s_{\bf q}}= \frac{1}{2}\left[s_{\bf q} \kappa_{\bf q} -s_{\bf p} \kappa_{\bf p}
\right]\left(\mathbf{h}^*_{s_{\bf p}}({\bf
p})\times\mathbf{h}^*_{s_{\bf q}}({\bf q})\right)\cdot
\mathbf{h}^*_{s_{\bf k}}({\bf k}).
\end{eqnarray}

\subsection{Resonant triads}

The helical mode decomposition~(\ref{waleffe1}) applies for any
velocity field, containing an arbitrary spectrum of wavevectors.
We restrict in the following the analysis to a set of three
interacting inertial waves of wavevectors (${\bf k},{\bf p},{\bf
q}$). Equation~(\ref{NavierStokes_waleffe}) shows that the
amplitude of the mode of wavevector ${\bf k}$ is related to the
two other modes ${\bf p}$ and~${\bf q}$ according to
\begin{eqnarray}
\left(\frac{\partial}{\partial t} +\nu\kappa^2 \right)A_{\bf k}=
C_{\bf k}A^*_{\bf p}A^*_{\bf q}e^{i(\sigma_{\bf k}+\sigma_{\bf
p}+\sigma_{\bf q})t} \label{NavierStokes_waleffe2}
\end{eqnarray}
where $C_{\bf k}$ is short-hand for $C_{\bf kpq}^{s_{\bf k}s_{\bf
p}s_{\bf q}}=C_{\bf kqp}^{s_{\bf k}s_{\bf q}s_{\bf p}}$. Cyclic
permutation of ${\bf k}$, ${\bf p}$ and ${\bf q}$ in
Eq.~(\ref{NavierStokes_waleffe2}) gives the two other relevant
interaction equations between the three waves. We further restrict
the analysis to plane inertial waves invariant along $y$ (i.e.,
$\mathbf{k}\cdot {\bf e_y = 0}$). The three considered helical
modes (\ref{waleffe2}) therefore reduce to
\begin{equation}\label{waleffe3}
\mathbf{h}_{s_{\bf r}}({\bf r})=\frac{m_{\bf r}\mathbf{e_x}-k_{\bf
r}\mathbf{e_z}}{\kappa_{\bf r}}- i s_{\bf r} \mathbf{e_y},
\end{equation}
where ${\bf r}$ stands for ${\bf k}$, ${\bf p}$ or ${\bf q}$. From
Eq.~(\ref{waleffe3}), the interaction coefficients
(\ref{waleffe5}) can be explicitly computed,
\begin{eqnarray}\label{waleffe6}
C_{\bf k}=\frac{i}{2 \kappa_{\bf k} \kappa_{\bf p} \kappa_{\bf q}}&\left[m_{\bf p} k_{\bf q}
-m_{\bf q}k_{\bf p}\right] [\kappa_{\bf q}^2 -\kappa_{\bf p}^2 + s_{\bf q} s_{\bf k} \kappa_{\bf q}\kappa_{\bf k} -s_{\bf p}
s_{\bf k}\kappa_{\bf p}\kappa_{\bf k} ],
\end{eqnarray}
and similarly for the two cyclic permutations.

Since in Eq.~(\ref{NavierStokes_waleffe2}) and in its two cyclic
permutations the $A_{\bf r}(t)$ coefficients have to be understood
as complex velocity amplitudes evolving slowly compared to wave
periods $2\pi/\sigma_{\bf r}$, temporal resonance is needed in
addition to spatial resonance for the left-hand
coefficients~$A_{\bf r}$ to be nonzero. Using $0,1,2$ for
reindexing the three waves ${\bf k}$, ${\bf p}$ and ${\bf q}$,
this leads to the triadic resonance conditions
\begin{eqnarray}
{\bf k_0 + k_1 + k_2 = 0}, \label{eq_sum_k} \\
\sigma_0 + \sigma_1 + \sigma_2 =0. \label{eq_sum_sigma}
\end{eqnarray}

\begin{figure}
    \centerline{\includegraphics[width=12cm]{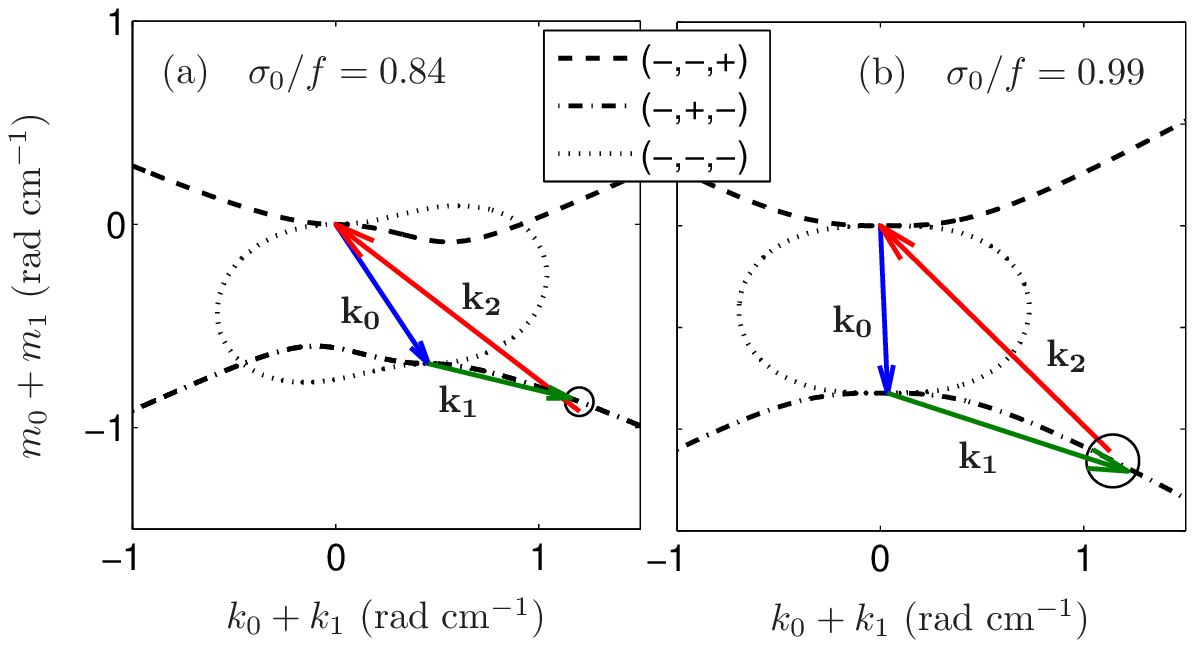}}
    \caption{(Color online) Resonance curves for the primary waves (a)
    [$s_0=-1, \sigma_0=0.84 f$, $\kappa_0=0.82$~rad~cm$^{-1}$] and (b)
    [$s_0=-1, \sigma_0=0.99 f$, $\kappa_0=0.82$~rad~cm$^{-1}$]. The
    curves represent the location of ${\bf k}_0+{\bf
    k}_1=(k_0+k_1,m_0+m_1)$ satisfying Eq.~(\ref{equation_k1m1}) for
    the 3 possible combinations of signs. The wavevectors measured
    experimentally are shown using arrows. The circle is the
    theoretical prediction for the location of ${\bf k}_0+{\bf k}_1$
    obtained from the maximum growth rate criterion, determined using
    the experimental primary wave amplitude [$A_0=0.29\pm
    0.07$~cm~s$^{-1}$ for (a) and $A_0=0.34\pm 0.11$~cm~s$^{-1}$ for
    (b)]. The diameter of the circle measures the uncertainty of the
    prediction due to the uncertainty on the wave amplitude $A_0$.
    \label{triad_exp12345}}
\end{figure}

We consider in the following that only the primary wave $A_0$, of
given frequency $\sigma_0$, wavevector ${\bf k_0}$ = $(k_0,m_0)$
and helicity sign $s_0$, is present initially in the system (i.e.,
$A_{1,2}(0)=0$). The two secondary waves ($s_1$, $\sigma_1$, ${\bf
k_1}$) and ($s_2$, $\sigma_2$, ${\bf k_2}$) which could form a
resonant triad with the primary wave may be determined using the
resonance conditions (\ref{eq_sum_k}) and (\ref{eq_sum_sigma}).
From the dispersion relation for inertial waves (\ref{disp_eq}),
the resonance conditions lead to
\begin{eqnarray}
s_0 \frac{m_0}{\sqrt{k_0^2+m_0^2}} & + & s_1 \frac{m_1}{\sqrt{k_1^2+m_1^2}} - s_2 \frac{m_0 + m_1}{\sqrt{(k_0+k_1)^2+(m_0+m_1)^2}}  =0.\label{equation_k1m1}
\end{eqnarray}
For a given primary wave $(s_0, k_0, m_0)$,  the solution of this
equation for each sign combination $(s_0,s_1, s_2)$ is a curve in
the $(k_1, m_1)$ plane (see Fig. \ref{triad_exp12345}). Without
loss of generality, once we have taken $s_0=-1$ (which corresponds
to the experimental configuration), it is necessary to consider
four sign combinations: $(-,-,-)$, $(-,+,-)$, $(-,-,+)$ and
$(-,+,+)$. Notice that the three first combinations always admit
solutions, whereas the fourth one, $(-,+,+)$, admits a solution
only if $|m_0|\leq\kappa_0/2$, i.e. $\theta>60$\degre. The
exchange of ${\bf k}_1$ and ${\bf k}_2$ keeps the $(-,-,-)$ and
$(-,+,+)$ resonances unchanged, but exchanges the $(-,-,+)$ and
$(-,+,-)$ resonances. Eventually, three independent sign
combinations remain: $(-,-,-)$, $(-,\mp,\pm)$ and $(-,+,+)$.

\subsection{Experimental verification of the resonance condition}

The predictions of the triadic resonance theory are compared here
with the measured wavevectors of the secondary waves.
Figure~\ref{triad_exp12345} shows the theoretical resonance curves
for two forcing frequencies, $\sigma_0/f=0.84$ and $0.99$. For
both curves, helicity sign and wavenumber of the primary wave are
chosen according to the experimental values,  $s_0=-1$ and
$\kappa_0=0.82$~rad~cm$^{-1}$.

For both frequencies $\sigma_0$ considered here, only the three
first sign combinations admit solutions. The $(-,-,-)$ combination
gives a closed loop, whereas the two others, $(-,\mp,\pm)$, give
infinite branches, tending asymptotically to constant angles. The
limit of large secondary wavevectors is such that
$|\sigma_1|=|\sigma_2|= |\sigma_0|/2$: when a wave ${\bf k}_0$
excites two waves of wavelength $\lambda\ll 2\pi/\kappa_0$, both
secondary waves have frequency $\sigma_0/2$, with opposite
wavevectors, leading to a stationary wave pattern. However, such
large wavenumbers are prevented by viscosity, as will be shown in
Sec.~\ref{sec:mgrc}.

Figure \ref{triad_exp12345} also shows the measured
secondary wavevectors ${\bf k}_1$ and ${\bf
k}_2$. These wavevectors are obtained from the
phase fields $\varphi_{1,2}$ extracted by Hilbert filtering, using
\begin{eqnarray}
{\bf k}_{1,2}={\bf \nabla} \varphi_{1,2}.
\label{wavevector}
\end{eqnarray}
These measurements are then averaged over regions of about
(130~mm)$^2$ where the secondary waves can be considered as
reasonably spatially monochromatic. It must be noted that a same
plane wave can be equivalently described by ($s$, $\sigma>0$,
${\bf k}$) and ($s$, $-\sigma<0$, ${-\bf k}$). Since we always
consider primary waves with positive frequency $\sigma_0>0$,
according to Eq.~(\ref{eq_sum_sigma}), the subharmonic frequencies
$\sigma_{1,2}$ have to be taken negative. As a consequence, the
Hilbert filtering should be performed for the negative peaks in
the temporal Fourier transform, in order to produce phase fields
with the appropriate sign. Practically, the Hilbert filtering has
been performed around the positive peaks $-\sigma_{1,2}$, and the
signs of the measured wavevectors have been changed accordingly.

The secondary wavevectors ${\bf k}_1=(k_1, m_1)$ and ${\bf
k}_2=(k_2, m_2)$ measured experimentally, shown in
Fig.~\ref{triad_exp12345}, are in good agreement with the triadic
condition (\ref{eq_sum_k}), forming a triangle such that ${\bf
k}_0 + {\bf k}_1 + {\bf k}_2 = 0$. Moreover, the apex of the
triangle, at ${\bf k}_0 + {\bf k}_1$, falls onto one of the three
resonant curves. The selected resonant curve corresponds to the
sign combination $(-,+,-)$, in agreement with the observed
experimental helicities. We actually verify that $s_1=\sigma_1
\kappa_1/f m_1$ is positive ($\sigma_1<0$ and $m_1<0$) and that
$s_2=\sigma_2 \kappa_2/f m_2$ is negative ($\sigma_2<0$ and
$m_2>0$), confirming the $(-,+,-)$ nature of the experimental
resonance.

Interestingly, the shape of the triangle ${\bf k}_0 + {\bf k}_1 +
{\bf k}_2 = 0$ in Fig.~\ref{triad_exp12345} indicates that the
group velocity of the secondary wave ${\bf k}_1$ is oriented
towards the wavemaker. Indeed, we recall that, for a given
wavevector ${\bf k}$, the group velocity ${\bf c}_g$ is normal to
${\bf k}$, and the vertical projections of~${\bf c}_g$ and ${\bf
k}$ are oriented in the same direction if $\sigma>0$ and in
opposite directions if $\sigma<0$. Accordingly,
Fig.~\ref{triad_exp12345} shows that ${\bf c}_{g0}$ and ${\bf
c}_{g2}$ are oriented downward, pointing from the wavemaker
towards the bottom of the tank, whereas ${\bf c}_{g1}$ is oriented
upward, pointing towards the wavemaker. As a consequence, the
secondary wave ${\bf k}_1$ is fed by the primary wave, but
releases its energy back to the wavemaker.

For all the primary wave angles for which the instability is
observed, the secondary waves are systematically such that
$|\sigma_{1}|$ and $|\sigma_{2}|$ are lower than $|\sigma_0|$. The
dispersion relation hence yields secondary wavevectors ${\bf
k}_{1,2}$  more horizontal than ${\bf k}_0$, as illustrated in
Fig.~\ref{triad_exp12345}. This property, which actually follows
from the conservation of energy and helicity,\cite{Smith1999}
illustrates the natural tendency of rotating flows to transfer
energy towards slow quasi-two-dimensional modes. If the process is
repeated, as in rotating turbulence, the energy becomes eventually
concentrated on nearly horizontal wavevectors, corresponding to a
quasi-2D flow, with weak dependence along the rotation
axis.~\cite{Cambon2008,Lamriben2011b}

\section{Selection of the most unstable resonant triad}

\subsection{Maximum growth rate criterion}
\label{sec:mgrc}

In order to univocally predict the resonant secondary waves, a
supplementary condition must be added to
Eq.~(\ref{equation_k1m1}): we assume that the selected resonant
triad is the one with the largest growth rate. Going back to the
wave interaction equations~(\ref{NavierStokes_waleffe2})
associated to the temporal resonance condition
(\ref{eq_sum_sigma}), the amplitudes of the secondary waves are
governed by
\begin{eqnarray}
\frac{d A_1}{d t}= C_{1}A^*_0 A^*_2 - \nu \kappa_1^2 A_1,\label{syst1}\\
\frac{d A_2}{d t}= C_{2}A^*_0 A^*_1 - \nu \kappa_2^2 A_2,\label{syst2}
\end{eqnarray}
with $C_{1,2}$ given by Eq.~(\ref{waleffe6}) taking ${\bf k} =
{\bf k}_{1,2}$ (see also Appendix A in
Ref.~\onlinecite{Smith1999}). Solving this system with initial
conditions $A_{1,2}(0)=0$, and assuming that $A_0$ remains almost
constant at short time, lead to the solutions
\begin{eqnarray}
A_{1,2}(t)=B_{1,2}\,(e^{\gamma_+ t} - e^{\gamma_- t}),
\end{eqnarray}
where the growth rates $\gamma_\pm$ write
\begin{eqnarray}
\gamma_\pm = -\frac{\nu}{2}(\kappa_1^2 +
\kappa_2^2) \pm \sqrt{\frac{\nu^2}{4}(\kappa_1^2 -
\kappa_2^2)^2+C_1C_2|A_0|^2}.\quad \label{Eq_lambda}
\end{eqnarray}
In the following, we consider the primary wave amplitude as real
without loss of generality, so $|A_0| = A_0$.

The coefficient $\gamma_-$ is always negative, so the stability of
the system is governed by the sign of $\gamma_+$, which we simply
note $\gamma$ in the following. Interestingly, this growth rate
$\gamma$ depends on the amplitude $A_0$ of the primary wave. As a
consequence, the primary wave is unstable with respect to a given
set of secondary waves, selected by the resonance condition and
unequivocally denoted by $\kappa_1$, only if $A_0$ exceeds the
threshold $A_c(\kappa_1) = \nu \kappa_1 \kappa_2/\sqrt{C_1C_2}$ in
which case $\gamma(\kappa_1)>0$. In other words, for a given
couple of secondary waves (denoted by $\kappa_1$) to be possibly
growing, the Reynolds number based on the primary wave, $Re_0 =
A_0/(\kappa_0 \nu)$, must exceed a critical value
$Re_c(\kappa_1)=A_c(\kappa_1)/(\kappa_0 \nu)$ for the onset of the
parametric instability. This critical Reynolds number is actually
an increasing function of $\kappa_1$ and tends to zero as
$\kappa_1 \rightarrow 0$, showing that whatever the value of
$Re_0$, there is always a continuum of resonant triads with
$Re_0>Re_c(\kappa_1)$, i.e. with a positive growth rate. The main
consequence is that, whatever the value of $Re_0$, the most
unstable triad always has a positive (maximum) growth rate and the
parametric instability does not have any $Re_0$ threshold to
proceed.

If viscosity can be neglected, Eq.~(\ref{Eq_lambda}) reduces to
$\gamma=\sqrt{C_1C_2}\,A_0$. In the limit of large secondary
wavenumbers $\kappa_{1,2}\gg \kappa_0$, one has ${\bf k}_1 \simeq
-{\bf k}_2$, and the growth rate $\gamma$ is found to tend
asymptotically toward a maximum value,\cite{Koudella2006} i.e.,
the selected secondary waves have frequency exactly half the
forcing frequency. Taking viscosity into account reduces the
growth rate of the large wavenumbers, and hence selects finite
wavenumbers. Equation~(\ref{Eq_lambda}) indicates that larger
wavenumbers are selected for larger primary wave amplitudes $A_0$
and/or lower viscosity, i.e. for larger Reynolds number $Re_0$.

\subsection{Selection of the most unstable wavenumbers}
\label{sec:smuw}

\begin{figure}
    \centerline{\includegraphics[width=6.5cm]{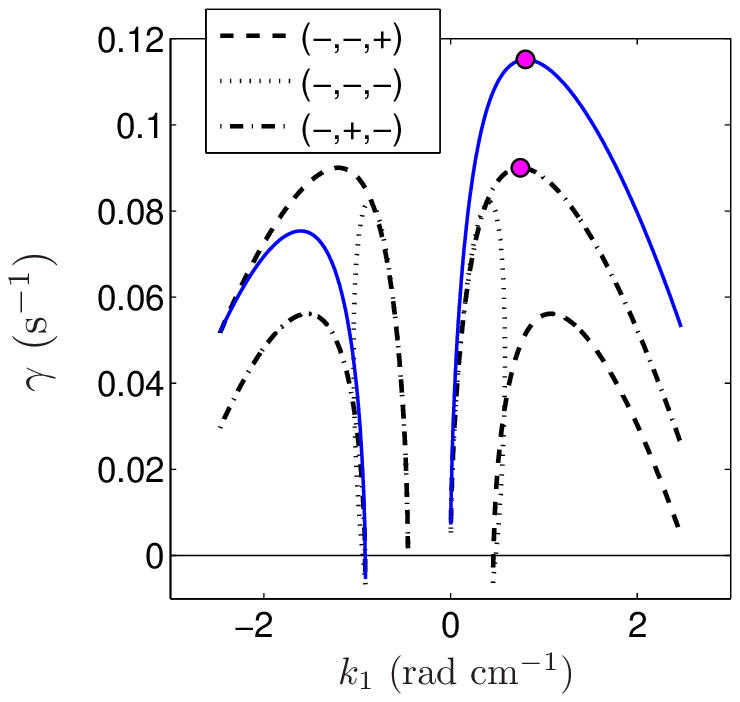}}
\caption{(Color online) Growth rates $\gamma$ as a function of
$k_1$, computed from Eq.~(\ref{Eq_lambda}), for the three possible
resonances for a primary wave $(s_0=-1, \sigma_0=0.84 f$,
$\kappa_0=0.82$~rad~cm$^{-1}$). The growth rates have been
computed using the average value $A_0=0.29$~cm~s$^{-1}$ for the
primary wave amplitude. For resonance $(-,+,-)$, an additional
curve (continuous line) has been computed using a wave amplitude
25\% larger.\label{lambda_k1}}
\end{figure}

In Fig.~\ref{lambda_k1}, the predicted growth rates $\gamma$ are
plotted  for the three possible sign combinations, for the primary
wave defined by $s_0=-1, \sigma_0=0.84 f$,
$\kappa_0=0.82$~rad~cm$^{-1}$. These growth rates have been
computed using the primary wave amplitude averaged over the area
where the secondary wavevectors have been measured (see the square
in Fig.~\ref{sub12}(a)), $A_0=0.29$~cm~s$^{-1}$. For the 3 types
of resonance, the growth rates tend to zero when $k_1 \rightarrow
-k_0/2$ and $k_1 \rightarrow \infty$ (because of viscosity). If
the secondary waves ${\bf k}_1$ and ${\bf k}_2$ are exchanged,
which amounts to exchange the $(-,-,+)$ and $(-,+,-)$ resonances,
the same growth rates are obtained: the curves for $(-,-,+)$ and
$(-,+,-)$ are symmetrical with respect to $k_0/2$.

Interestingly, the growth rate is positive for a broad range of
wavenumbers. Together with the broad subharmonic peaks observed in
the temporal spectrum of Fig.~\ref{spectre}, this confirms that
the parametric resonance is weakly selective in this system.
Values of $k_1$ corresponding to significant growth rates are of
the same order of magnitude as the primary wavenumber
$\kappa_0=0.82$~rad~cm$^{-1}$, indicating that the viscosity has a
significant effect on the selection of the excited resonant triad.
For the value of $\sigma_0/f$ considered in Fig.~\ref{lambda_k1},
the maximum growth rate is obtained for the $(-,+,-)$ resonance,
for $k_1^{\rm max}=0.75$~rad~cm$^{-1}$. The corresponding
predicted wavevector ${\bf k}_1$ is represented as a circle in the
resonance curve of Fig.~\ref{triad_exp12345}(a), and is found in
excellent agreement with the experimental measurement of ${\bf
k}_1$ (shown with an arrow).

Because of the viscous attenuation, the primary wave amplitude
$A_0$ actually depends on the distance from the wavemaker. In the
measurement area shown in Fig.~\ref{sub12}(a), spatial variations
of $\pm 25\%$ are found around the average $A_0=0.29$~cm~s$^{-1}$.
Since the growth rate (\ref{Eq_lambda}) depends on $A_0$, this
introduces an uncertainty on the predicted value of $\gamma$, and
consequently on the selected secondary wavenumbers. In order to
appreciate the influence of the measured value of $A_0$ on the
predicted triadic resonance, we also plot in Fig.~\ref{lambda_k1}
the growth rate of the selected $(-,+,-)$ resonance, but for a
value of $A_0$ increased by an amount of 25\% (continuous line),
which corresponds to the wave amplitude in the close vicinity of
the wavemaker. The maximum growth rate is actually found to
strongly depend on $A_0$, with an increase of 30\%, indicating
that the onset of the parametric instability will take place first
close to the wavemaker. This strong sensitivity would make any
direct comparison with an experimental growth rate too difficult.
On the other hand, the selected wavenumber $k_1^{\rm max}$ is
quite robust, showing a slight increase of 6\% only when $A_0$ is
increased by 25\%. As a consequence, the uncertainty in the
measurement of $A_0$, which is unavoidable because of the viscous
attenuation of the primary wave, does not affect significantly the
prediction for the most unstable secondary wavevectors.

The size of the circles in Figs.~\ref{triad_exp12345}(a) and (b)
illustrates the uncertainty in the determination of the most
unstable wavevectors due to the spatial variation of $A_0$. The
relative uncertainty lies in the range $5 - 15\%$ for the range of
wave frequencies considered here. In spite of this uncertainty, we
can conclude that the secondary wavevectors predictions from the
maximum growth rate criterion are in good agreement with the
observed resonant triads.

\subsection{Dependence of the secondary waves properties on the primary wave frequency}

\begin{figure}
    \centerline{\includegraphics[width=15cm]{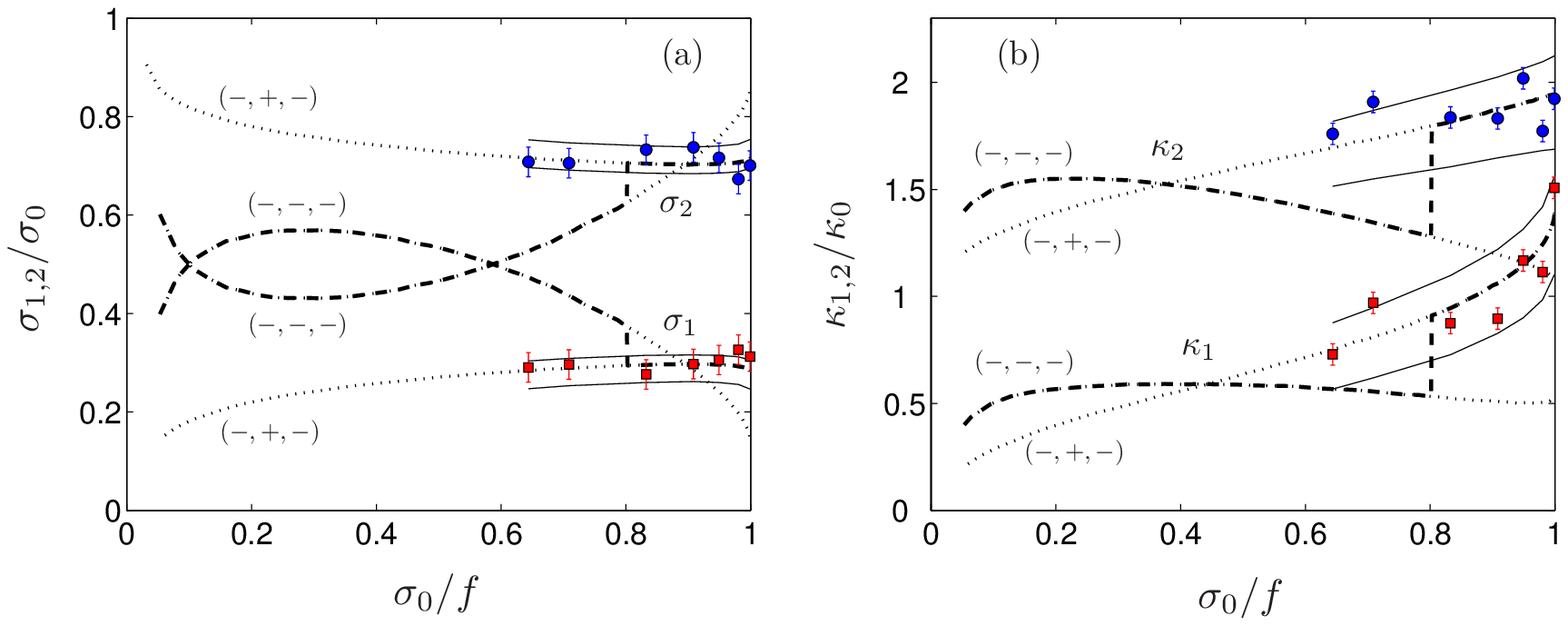}}
    \caption{(Color online) Normalized frequencies $\sigma_{1,2} /
    \sigma_0$ (a), and wavenumbers $\kappa_{1,2}/\kappa_0$ (b) of the
    secondary waves,  as a function of the primary wave frequency
    $\sigma_0/f$. Filled circles and squares with errorbars correspond
    to experimental measurements. Predictions from the triadic
    resonance instability are represented with dashed thick lines
    (using absolute maximum growth rate criterion) and dotted lines
    (using maximum growth rate criterion for the ($-,-,-$) and
    ($-,+,-$) resonances). Predictions for the most unstable resonance
    are ($-,-,-$) for $\sigma_0/f<0.79$, and ($-,+,-$) for
    $\sigma_0/f>0.79$. These predictions have been computed with a
    typical amplitude $A_0=0.30$~cm~s$^{-1}$ for the primary wave.
    Continuous solid lines show the allowed range around the ($-,+,-$)
    curves, determined by considering an uncertainty of $\pm 50\%$ on
    $A_0$.} \label{freq}
\end{figure}

We finally characterize here the evolution of the secondary wave
properties (frequencies and wavenumbers) as the frequency of the
primary wave is changed. For a given primary wave amplitude $A_0$,
the secondary frequencies $\sigma_{1,2}$ and wavenumbers
$\kappa_{1,2}$ have been systematically computed  according to the
maximum growth rate criterion, and are reported in Fig.~\ref{freq}
as a function of $\sigma_0 / f \in [0,1]$. The dotted lines
correspond to the $(-,-,-)$ and $(-,+,-)$ resonances, whereas the
dashed thick lines are computed from the absolute maximum growth
rate among all the possible resonances. For $\sigma_0/f>0.79$, the
growth rate is maximum on the $(-,\pm,\mp)$ branch, whereas for
$\sigma_0/f<0.79$ it is maximum on the $(-,-,-)$ branch.

In Fig.~\ref{freq}, we also show the experimental measurements of
$\sigma_{1,2}$ and $\kappa_{1,2}$ for the range of primary wave
frequencies for which a subharmonic instability is observed,
$0.65<\sigma_0/f<0.99$. The errorbars show the uncertainties
computed from the measured frequencies and wavenumbers. The
agreement with the predictions from the triadic resonance theory
is excellent for the $(-,+,-)$ branch. However, it is not clear
why all the measurements actually follow the $(-,+,-)$ branch,
although the $(-,-,-)$ branch is expected to be more unstable for
the two data points at $\sigma_0/f < 0.79$.

The limited spatial extent of the primary wave along its
transverse direction (which represents 4 wavelengths only) and its
amplitude decay along its propagation direction (because of
viscous attenuation) may be responsible for this unexpected
stability of the $(-,-,-)$ branch at low $\sigma_0/f$. Indeed, the
$(-,-,-)$ branch is associated to wavelengths significantly larger
than the primary wavelength, so that a large spatial region of
nearly homogeneous primary wave amplitude is required to sustain
such large wavelength secondary waves. On the other hand, the
$(-,+,-)$ resonance generates lower wavelengths, which can more
easily fit into the limited extent of the primary wave. Finite
size effects may therefore explain both the preferred $(-,+,-)$
resonance at $\sigma_0/f < 0.79$, and the unexpected global
stability of the primary wave for $\sigma_0/f < 0.65$. Confinement
effects are not described by the present triadic resonance theory,
which assumes plane waves of infinite spatial extent. Apart from
this open issue, we can conclude that, at least for sufficiently
large forcing frequency, the observed secondary frequencies and
wavenumbers are in good quantitative agreement with the
predictions from the triadic resonance theory.

\section{Discussion and conclusion}

Using a wavemaker initially designed to generate beams of {\em
internal} gravity waves in stratified fluids, we have successfully
generated well-defined plane {\em inertial} waves in a rotating
water tank. Spectral analysis, performed on particle image
velocimetry measurements of this plane inertial wave, has revealed
the onset of a parametric instability, leading to the emergence of
two secondary subharmonic waves. The wavevectors and frequencies
of the primary and secondary waves are found in good agreement
with the spatial and temporal resonance conditions for a resonant
triad of inertial waves. Moreover, using the triadic resonance
theory for inertial waves derived by Smith and
Waleffe,~\cite{Smith1999} the growth rate of the instability has
been computed, yielding predictions for the secondary wavevectors
and frequencies in agreement with the measurements. At low forcing
frequency, we observe a departure from these predictions which may
be associated to the finite size of the primary wave. These finite
size effects cannot actually be described within the triadic
resonant theory, which relies on plane waves of infinite extent.

Triadic resonant instability for inertial and internal waves share
a number of common properties. In particular, equations governing
the wave amplitudes equivalent to Eqs.~(\ref{syst1}) and
(\ref{syst2}) may also be derived for a triad of internal waves,
but in this case, they concern the amplitude of streamfunctions
and not of velocities.~\cite{Koudella2006} The interaction
coefficients for internal waves $\tilde{C}_{r}$ (with $r=0,1,2$)
can be readily obtained from the interaction coefficients for
inertial waves $C_{r}$ through a simple exchange of the vertical
and horizontal components of the wavevectors, and introducing a
prefactor:
\begin{equation}
\tilde{C}_{r}(k,m) = \frac{\kappa_p \kappa_q}{\kappa_r}
C_{r}(m,k).
\end{equation}
The $\kappa_p \kappa_q/\kappa_r$ prefactor between the two types
of coefficients comes from the fact the wave amplitude is directly
given by the velocity ${\bf u}$ in the case of inertial waves,
whereas it is given by the streamfunction $\widetilde{\psi} \sim
u/\kappa$ in the case of internal waves. The exchange of the
vertical and horizontal components of the wavevectors comes from
the comparison between the dispersion relations for inertial and
internal waves, $\sigma/f=s m/\kappa$ and $\sigma/N=s k/\kappa$
respectively, with $f=2\Omega$ the Coriolis parameter and $N$ the
Brunt--V\"{a}is\"{a}l\"{a} frequency. The inviscid growth rate of
the parametric instability $\tilde \gamma$ for the internal waves
is actually equal to the one of inertial waves $\gamma$ through
\begin{equation}
\tilde \gamma= \sqrt{\tilde{C}_{1}\tilde{C}_{2}}\widetilde{A}_0 = \sqrt{C_{1}C_{2}} \kappa_0 \widetilde{A}_0 = \gamma,
\end{equation}
where $\widetilde{A}_0$ is the primary internal wave amplitude
(homogeneous to a streamfunction). Here, the inertial wave
amplitude $A_0$ (homogeneous to a velocity) identifies with
$\kappa_0 \widetilde{A}_0$. This equality between inertial and
internal growth rates finally shows that the predicted secondary
waves should be identical for the two types of waves.

Interacting inertial waves are of primary importance for the
dynamics of rotating turbulence. In the limit of low Rossby
numbers $Ro = U / \Omega L$, where $U$ and $L$ are characteristic
velocity and length scales, rotating turbulence can be described
as a superposition of weakly interacting inertial waves, whose
interactions are directly governed by triadic resonances. This is
precisely the framework of wave turbulence as analyzed in
Refs.~\onlinecite{Galtier03} and \onlinecite{Cambon04} in the
context of rotating turbulence. The parametric instability between
three inertial waves can be seen as an elementary process by which
energy is transferred between wavevectors in rotating turbulence.
This anisotropic energy transfer takes place both in scales (or
wavenumbers) and directions (or angles). The {\it angular} energy
transfer is always directed towards more horizontal wavevectors,
providing a clear mechanism by which slow quasi-2D motions become
excited.\cite{Smith1999} However, the nature of energy transfers
through triadic resonance in terms of {\it wavenumbers} (or {\it
scales}) ---i.e., whether the energy proceeds from large to small
scales or inversely--- is found to depend on wave amplitude and
viscosity. Indeed, it can be shown theoretically, within the
present triadic resonance framework, that waves of amplitude large
compared to $\nu \kappa_0$ are unstable with respect to secondary
waves of large wavenumbers, producing a direct energy cascade
towards small scales. On the other hand, waves of amplitude much
lower than $\nu \kappa_0$ are found to excite secondary waves of
smaller wavenumber, hence producing an inverse energy cascade
towards larger scales. The net result of this competition is
delicate to decide, and may contain an answer to the debated issue
concerning the direction of the energy cascade in rapidly rotating
turbulence.

\acknowledgments

We thank M. Moulin for the technical work and improvement made on
the wavemaker, and C. Borget for experimental help with the
rotating platform. The collaboration between FAST laboratory and ENS
Lyon Physics laboratory is funded by the ANR grant no.
ANR-2011-BS04-006-01 ``ONLITUR''. The rotating platform
``Gyroflow'' was funded by the ANR grant no. 06-BLAN-0363-01
``HiSpeedPIV'' and by the ``Triangle de la Physique''. ENS Lyon's
research work has been also partially supported by the ANR grant
no. ANR-08-BLAN-0113-01 ``PIWO''.


\begin{thebibliography}{}

\bibitem{Greenspan1968}
H. Greenspan, ``The Theory of Rotating Fluids,'' (Cambridge
University Press, London, 1968).

\bibitem{Lighthill1978}
J. Lighthill, ``Waves in Fluids,'' (Cambridge University Press,
London, 1978).

\bibitem{Pedlosky1987}
J. Pedlosky,  ``Geophysical Fluid Dynamics,'' (Springer-Verlag,
Heidelberg, 1987).

\bibitem{Phillips1963}
O.M. Phillips,  ``Energy transfer in rotating fluids by reflection
of inertial waves,'' Phys. Fluids {\bf 6}, 513 (1963).

\bibitem{Gostiaux2006}
L. Gostiaux, T. Dauxois, H. Didelle, J. Sommeria and S. Viboud,
``Quantitative laboratory observations of internal wave reflection
on ascending slopes,'' Phys. Fluids {\bf 18}, 056602 (2006).

\bibitem{Fultz1959}
D. Fultz,  ``A note on overstability and the elastoid-inertia
oscillations of Kelvin, Soldberg, and Bjerknes,'' J. Meteo. {\bf
16}, 199--207 (1959).

\bibitem{McEwan1970}
A. D. McEwan,  ``Inertial oscillations in a rotating fluid
cylinder,'' J. Fluid Mech. {\bf 40}, 603--639 (1970).

\bibitem{Manasseh1994}
R. Manasseh,  ``Distortions of inertia waves in a rotating fluid
cylinder forced near its fundamental mode resonance,'' J. Fluid
Mech. {\bf 265}, 345--370 (1994).

\bibitem{Maas2001}
L. R. M. Maas,  ``Wave focusing and ensuing mean flow due to
symmetry breaking in rotating fluids,'' J. Fluid Mech. {\bf 437},
13--28 (2001).

\bibitem{Duguet2005}
Y. Duguet, J. F. Scott, L. Le Penven, ``Instability inside a rotating
gas cylinder subject to axial periodic strain,''
Phys. Fluids {\bf 17}, 114103 (2005).

\bibitem{Duguet2006}
Y. Duguet, J. F. Scott, L. Le Penven, ``Oscillatory jets and
instabilities in a rotating cylinder,''
Phys. Fluids {\bf 18}, 104104 (2006).

\bibitem{Meunier2008}
P. Meunier, C. Eloy, R. Lagrange and F. Nadal,  ``A rotating fluid
cylinder subject to weak precession,'' J. Fluid Mech. {\bf 599},
405--440 (2008).

\bibitem{Bewley2007}
G. P. Bewley, D. P. Lathrop, L. R. M. Maas, and K. R. Sreenivasan,
``Inertial waves in rotating grid turbulence,'' Phys. Fluids {\bf
19}, 071701 (2007).

\bibitem{Lamriben2011a}
C. Lamriben, P.-P. Cortet, F. Moisy and L.R.M. Maas, ``Excitation
of inertial modes in a closed grid turbulence experiment under
rotation,'' Phys. Fluids {\bf 23}, 015102 (2011).

\bibitem{Messio2008}
L. Messio, C. Morize, M. Rabaud and F. Moisy, ``Experimental
observation using particle image velocimetry of inertial waves in
a rotating fluid,'' Exp. Fluids {\bf 44}, 519-528 (2008).

\bibitem{Cortet2010}
P.-P. Cortet, C. Lamriben and F. Moisy, ``Viscous spreading of an
inertial wave beam in a rotating fluid,'' Phys. Fluids {\bf 22},
086603 (2010).

\bibitem{Thorpe1969}
S. A. Thorpe, ``On standing internal gravity waves of finite
amplitude,'' J. Fluid Mech. {\bf 32}, 489 (1969).

\bibitem{McEwan1971}
A. D. McEwan, ``Degeneration of resonantly-excited standing
internal gravity waves,'' J. Fluid Mech. {\bf 50}, 431 (1971).

\bibitem{Benielli1998}
D. Benielli and J. Sommeria, ``Excitation and breaking of
internal gravity waves by parametric instability,'' J. Fluid Mech.
{\bf 374}, 117 (1998).

\bibitem{Staquet2002}
C. Staquet and J. Sommeria, ``Internal gravity waves: From
instabilities to turbulence,'' Ann. Rev. Fluid Mech. {\bf 34},
559-593 (2002).

\bibitem{Olbers1981}
D. J. Olbers and N. Pomphrey, ``Disqualifying 2 candidates for the
energy-balance of oceanic internal waves,'' J. Phys. Ocean. {\bf
11}, 1423 (1981).

\bibitem{Kunze2004}
E. Kunze and S. G. Llewellyn Smith, ``The Role of Small-Scale
Topography in Turbulent Mixing of the Global Ocean,'' Oceanography
{\bf 17}, 55 (2004).

\bibitem{MacKinnon2005}
J. A. MacKinnon and K. B. Winters, ``Subtropical catastrophe:
Significant loss of low-mode tidal energy at 28.9\degre,''
Geophys. Res. Lett. {\bf 32}, L15605 (2005).

\bibitem{Koudella2006}
C. R. Koudella and C. Staquet, ``Instability mechanisms of a
two-dimensional progressive internal gravity wave,'' J. Fluid
Mech. {\bf 548}, 165--196 (2006).

\bibitem{Smith1999}
L. M. Smith and F. Waleffe, ``Transfer of energy to two-dimensional
large scales in forced, rotating, three-dimensional turbulence,''
Phys. Fluids {\bf 11}, 1608 (1999).

\bibitem{Cambon2008}
P. Sagaut and C. Cambon, ``Homogeneous turbulence dynamics,'' Cambridge
(2008).

\bibitem{Lamriben2011b}
C. Lamriben, P.-P. Cortet and F. Moisy, ``Direct Measurements of
Anisotropic Energy Transfers in a Rotating Turbulence
Experiment,'' Phys. Rev. Lett. {\bf 107}, 024503 (2011).

\bibitem{Staplehurst2008}
P.J. Staplehurst, P.A. Davidson, S.B. Dalziel, ``Structure
formation in homogeneous freely decaying rotating turbulence,'' J.
Fluid Mech. {\bf 598}, 81-105 (2008).

\bibitem{Gostiaux2007}
L. Gostiaux, H. Didelle, S. Mercier and T. Dauxois, ``A novel
internal waves generator,'' Exp. Fluids {\bf 42}, 123--130 (2007).

\bibitem{Mercier2008}
M. Mercier, N. Garnier and T. Dauxois, ``Reflection and
diffraction of internal waves analyzed with the Hilbert
transform,'' Phys. Fluids {\bf 20}, 0866015 (2008).

\bibitem{Mercier2010}
M. Mercier, D. Martinand, M. Mathur, L. Gostiaux, T. Peacock and
T. Dauxois, ``New wave generation,'' J. Fluid Mech. {\bf 657},
310--334 (2010).

\bibitem{Davis}
DaVis, LaVision GmbH, Anna-Vandenhoeck-Ring 19, 37081 Goettingen,
Germany.

\bibitem{pivmat}
F. Moisy, PIVMat toolbox for Matlab, {\text
http://www.fast.u-psud.fr/pivmat}.

\bibitem{Croquette1989}
V. Croquette and H. Williams, ``Nonlinear waves of the oscillatory
instability on finite convective rolls,'' Physica D {\bf 37},
300-314 (1989).

\bibitem{Waleffe1992}
F. Waleffe, ``The nature of triad interactions in
homogeneous turbulence,''
Phys. Fluids A {\bf 4} (2), 350 (1992).

\bibitem{Waleffe1993}
F. Waleffe, ``Inertial transfers in the helical decomposition,''
Phys. Fluids A {\bf 5} (3), 577 (1993).

\bibitem{fn_sign}
The definition of the helical mode used here corresponds to the
one in Ref.~\onlinecite{Smith1999}. In
Refs.~\onlinecite{Waleffe1992} and \onlinecite{Waleffe1993}, the
helical mode is defined as the complex conjugate of
Eq.~(\ref{waleffe2}), resulting in a sign change of the helicity.

\bibitem{Galtier03}
S. Galtier, ``Weak inertial-wave turbulence theory,'' Phys. Rev. E
{\bf 68}, 015301(R) (2003).

\bibitem{Cambon04} C. Cambon, R. Rubinstein, and F. S. Godeferd, ``Advances in wave
turbulence: rapidly rotating flows,'' New J. Phys. {\bf 6}, 73
(2004).

\end{thebibliography}
\end{document}